\def \yskip{\penalty-50\vskip3pt plus 3pt minus 2pt}
\def \reference{\par \yskip \noindent \hangindent .4in \hangafter 1}
\def \abc#1#2#3#4 {\reference#1, {\sl#2}, {\bf#3}, #4}
\def \blank {\lower 5pt\hbox to 0.75in{\hrulefill}}

\def \cm{~\rm{cm}}
\def \s{~\rm{s}}
\def \km{~\rm{km}}

\def \K{~\rm{K}}
\def \g{~\rm{g}}

\def \AU{~\rm{AU}}
\def \erg{~\rm{erg}}

\def \yr{~\rm{yr}}

\def \lesssim{\mathrel{<\kern-1.0em\lower0.9ex\hbox{$\sim$}}}
\def \gtrsim{\mathrel{>\kern-1.0em\lower0.9ex\hbox{$\sim$}}}

\documentstyle[11pt,aasms,tighten,flushrt]{article}

\begin{document}
\small

\setcounter{page}{1}



\title{MAGNETIC ACTIVITY OF THE COOL COMPANION
IN SYMBIOTIC SYSTEMS} 

\author{Noam Soker}

\affil{Department of Physics, Oranim, Tivon 36006, Israel; 
soker@physics.technion.ac.il}


$$
$$

\centerline {\bf ABSTRACT}

I argue that cool giant companions in most symbiotic binary systems
possess magnetic activity on a much higher level than
isolated, or in wide binary systems, cool giants.
Based on the behavior of main sequence stars, I assume that
magnetic activity and X-ray luminosity increase
with rotation velocity.
I then show that the cool companions in symbiotic systems are likely
to rotate much faster than isolated, or in wide binary systems,
cool giants.
The magnetic activity of the cool giant may be observed as
a global axisymmetrical mass loss geometry from the cool giant
(before the hot companion influences the outflow),
a stochastic mass loss process, i.e., variation of mass loss rate
with time and location on the giant's surface, and in relatively
strong X-ray emission.
The variation in the mass loss process from the cool giant may
cause variation in the properties of jets blown by the hot
compact companion.
I conclude that symbiotic systems should be high-priority X-ray targets.

{\it Subject headings:} binaries: symbiotic
$-$ circumstellar matter
$-$ stars: AGB
$-$ stars: magnetic fields

\section{INTRODUCTION} 

The level and the nature of the magnetic activity in red giant
branch (RGB) and asymptotic giant branch (AGB) stars is poorly
known and understood. 
Indications of magnetic activity come from polarized maser
emission and X-ray emission from a number of AGB stars.
Polarized SiO maser emission is observed close, typically less than a
few stellar radii, to some evolved stars (e.g. Kemball \& Diamond 1997),
whereas polarized OH maser emission is observed  at
$\sim 10^{15}-10^{16} \cm$ from the star (e.g., Zijlstra et al.\ 1989;
Szymczak, Cohen, \& Richards 1999; Miranda et al.\ 2001).
X-ray emission found in several evolved stars (e.g.,
H\"unsch et al.\ 1998; Schr\"oder, H\"unsch \& Schmitt 1998;
H\"unsch 2001) may indicate the presence of a hot
corona that results from magnetic activity.

The dynamo in evolved stars, e.g., AGB stars, is different in
many respects from the dynamo mechanism in the Sun.
The main difference is the ratio between the rotation
period, $P_{\rm rot}=2 \pi /\omega$, and the convective overturn
time $\tau_c$ (Soker \& Zoabi 2002).
This is defined as the Rossby number
${\rm Ro} \equiv (\omega \tau_c)^{-1}$.
 In the sun and other magnetically active main sequence stars,
$Ro<1$, whereas in evolved stars $Ro \gg 1$. 
In cool giant stars, therefore, the amplification is most likely
via a turbulent dynamo, i.e., an $\alpha\alpha$ dynamo,
but where the rotation plays a positive role
(Soker 2000b; Soker \& Zoabi 2002).
This mechanism is termed $\alpha^2 \omega$ dynamo.
  If indeed rotation plays a significant role in the amplification
of the magnetic field in cool giant stars, then systems where these
stars rotate fast would be expected to be more active.
Such systems are symbiotic stars (SSs), which are binary systems
composed of a mass losing evolved star, basically an RGB or an AGB
star, and an accreting compact star, in most cases a white dwarf (WD),
and in the minority of the systems a close main sequence (MS) star
(e.g., Belczynski, et al.\ 2000).

In $\S 2$ I show that the cool companion in symbiotic systems are
likely to rotate much faster than single cool giant stars, hence
they are prime candidates for possessing (relatively) strong
magnetic activity.
The presence of magnetic fields coronae in SSs were proposed in the past,
e.g., to account for UV emission (Stencel \& Sahade 1980).
A thorough review of binary evolution and interaction in symbiotic stars
is given by Iben \& Tutukov (1996).
I will only concentrate on the expected rotation speed of AGB
cool companions in these systems, and the expected magnetic activity.
In $\S 3$ I discuss the manifestation of the expected magnetic activity,
and in $\S 4$ I present my main conclusions, and
propose some future observations. 

\section{SPINNING-UP THE COOL COMPANION} 

\subsection{Accretion during the main sequence phase}
In this subsection I consider SSs with a WD hot companion.
I show that the presently cool giant companion was spun-up
when it was an MS star and the presently WD was an AGB star.
In most SSs the accretion rate into the WD companion is
$\dot M_{\rm acc} >10^{-9} M_\odot \yr^{-1}$, the orbital
period is $P_{\rm orb} < 10^4~$day, the mass loss rate 
(defined positively) from the cool companion is
$\dot M_w < 10^{-5} M_\odot \yr$, and the WD mass is
$M_{WD} \gtrsim 0.6 M_\odot$.
 The parameters of most systems, in particular the mass of the
WD companion, indicate that the progenitor of the WD
companions was relatively massive $M_p \gtrsim 3 M_\odot$.
Such stars lose $M_l \simeq 1-2 M_\odot$ as a slow wind during
their AGB phase.
Some fraction of this mass is accreted by the MS progenitor of
the presently giant star. 
To find the accreted angular momentum by the MS progenitor
of the present cool giant, I use the Bondi-Hoyle mass accretion
rate, and the angular momentum entering the Bondi-Hoyle
accretion cylinder (Wang 1981), i.e., having impact parameter
$b<R_a=2GM_g/v_r^2$, where $R_a$ is the accretion radius,
$M_g$ is the mass of the accreting star, and $v_r$ the relative
velocity of the star and the wind.
 The accreted angular momentum onto the MS progenitor of the
present giant is given by
(for $R_a<a$; see relevant equations and more references in Soker 2001) 
\begin{equation}
J_{\rm acc}=\frac{\eta}{4} \left( \frac{R_a}{a} \right)^4
\frac {M_l}{\mu} [G (M_g+M_p) a \mu^2 ]^{1/2},
\end{equation}
where $\mu = M_g M_p/(M_g+M_p)$ is the reduced mass of the
binary system, $M_p$ is the mass of the WD's progenitor, 
$\eta$ is the ratio of the accreted angular momentum to that
entering the Bondi-Hoyle accretion cylinder, and $a$ is the
orbital separation (for simplicity a circular orbit is assumed).
The last term in equation (1) (the $[]^{1/2}$ term) is the total
orbital angular momentum of the binary system.
 I took the wind velocity $v_w$ to be equal to
the relative velocity $v_r=(v_w^2+v_o)^{1/2}$, where $v_o$ is the
orbital velocity of the accreting star relative to the mass losing
star. This is justified for the parameters used here and the
other uncertainties, e.g., in the value of $\eta$. 
Scaling the different variables with typical values, I find the
accreted angular momentum to be
\begin{equation}
J_{\rm acc} \simeq 10 J_\odot
\left( \frac {\eta}{0.2} \right)
\left( \frac {M_g+M_p}{4 M_\odot} \right)^{1/2}
\left( \frac {M_g}{1.0 M_\odot} \right)^{4}
\left( \frac {M_l}{1 M R_\odot} \right)
\left( \frac {a}{50 \AU} \right)^{-7/2}
\left( \frac {v_r}{15 \km \s^{-1}} \right)^{-8} ,
\end{equation}
where $J_\odot=1.7 \times 10^{48} \g \cm^2 \s^{-1}$ is the present
angular momentum of the Sun.
Stars more massive than the sun rotate faster, up to
100 times faster, but accrete much more angular momentum, e.g.,
by a factor of $\sim 100$ for $M_g=3 M_\odot$.
Of course, the MS can't rotate faster than the break-up
velocity. 
The conclusion from the last equation is that the present giant
in most SSs reach the RGB and AGB with typical angular momentum,
hence angular velocity, much larger than isolated stars do.

As an example I consider Mira (o Ceti) a non-SS binary system with
an AGB star and a WD in a projected orbital separation of $\sim 70 \AU$ 
(Karovska et al.\ 1997). 
The true orbital separation is larger, but it was smaller before mass
was lost by the two components.
 Taking therefore $a=70 \AU$ in equation (2) with all other parameters
held the same, gives $J_{\rm acc} \simeq 3 J_\odot$. 
If, on the other hand, the slow wind speed were lower, as is
quite plausible for a large AGB star, as the progenitor of the WD was,
say $v_w=10 \km \s^{-1}$, then $J_{\rm acc} \simeq 75 J_\odot$.
I suggest, therefore, that Mira has reached the AGB with an angular
momentum larger than typical isolated AGB stars do, despite the
orbital separation that is too large to allow substantial tidal
spin-up (see next subsection).
Such an extra spin may enhance magnetic activity (see $\S 3$),
which may lead to a stochastic mass loss process ($\S 3.2$), which
may explain Mira's asymmetrical shape observed by Karovska et al.\ (1997)
and Marengo et al.\ (2001).

\subsection{Tidal interaction}

The hot companion spins-up the cool giant companion via tidal interaction,
whereas mass loss from the giant spins it down.
I use the equilibrium tidal interaction for giants (Zahn 1977; 1989;
Verbunt \& Phinney 1995) with time scales from Soker (1998).
The circularization time of the orbit is given by
\begin{eqnarray}
\tau_{\rm {circ}} =  
1.2 \times 10^8
{\frac{1}{f_s}} 
\left( {\frac{L_g}{2000 L_\odot}} \right)^ {-1/3}
\left( {\frac{R_g}{200 R_\odot}} \right)^ {2/3}
\left( {\frac{M_{\rm {env}}}{M_g}} \right)^ {-1} 
\left( {\frac{M_{\rm {env}}}{M_\odot}} \right)^ {1/3} \nonumber \\
\times \left( {\frac{M_2}{M_g}} \right)^ {-1}
\left( 1+ {\frac{M_2}{M_g}} \right)^ {-1}
\left( \frac{a}{10R_g} \right)^ {8}
\yr ,
\end{eqnarray}
where $L_g$, $R_g$ and $M_g$ are the luminosity, radius, and total mass
of the giant star, $M_{\rm {env}}$ is the giant's envelope mass,
$M_2$ is the mass of companion that exerts the tide, $a$ is the
semi-major axis, and $f_s(e)$ is a steeply increasing
function of eccentricity $e$ (Hut 1982) with $f_s(0) \simeq 1$.
The transferring rate of orbital angular momentum to giant spin is
\begin{equation}
\dot J_{\rm orb} \simeq \frac{\dot a}{2a} J_{\rm orb} \simeq
\frac {J_{\rm orb}}{\tau_{\rm cir}}, 
\end{equation}
where $J_{\rm orb}$ is the total orbital angular momentum of the
two stars, and dot stands for time derivative.
 The expression is accurate for a circular orbit; for eccentric
orbits there is a factor of correction (see Hut 1982 for the
dependence of the different factors on eccentricity).
 I don't consider this factor here, because the mass loss process,
if it depends on the orbital phase, may lead to a change in
eccentricity, which is not considered here either.
 The giant slows-down because of angular momentum loss
via mass loss, at a rate given by
\begin{equation}
\dot J_w = \beta R_g^2 \omega \dot M_w,
\end{equation}
 where $\omega$ is the giant's angular velocity,
$\dot M_w$ is the mass loss rate in the wind, and $\beta$ is a
parameter that depends on the mass loss geometry;
$\beta=2/3$ for spherical mass loss and $\beta=1$ for
equatorial mass loss.
 Both  $J_w$ and $M_w$ are defined positively.
In systems in which the evolution time is less than the
circularization time, the angular velocity of the giant is given
by equating the rate in equation (4) to that in equation (5), where
I neglect the nagular momentum of the spining giant's envelope.
This gives 
\begin{equation}
\frac{\omega}{\omega_{\rm Kep}}
\simeq \left( \frac{M_g+M_2}{M_g} \right)^{1/2}
\left( \frac{a}{R_g} \right)^{1/2}
\frac{\mu}{\dot M_w}
\frac{(1-e^2)^{1/2}}{\tau_{\rm cir}}
\frac{1}{\beta},
\end{equation}
where $\omega_{\rm Kep}$ is the Keplerian angular velocity on
the equator of the giant, and $\mu=M_gM_2/(M_g+M_2)$ is the
reduced mass of the binary system.
Substituting typical values and omitting terms of weak dependence on
several factors in equation (3) (in any case, as indicated
above, $\tau_{\rm cir}$ serves as approximation for the rate
of angular momentum transfer in eccentric orbits) and
taking $M_2 \simeq M_g \simeq M_\odot$, this ratio is
\begin{eqnarray}
\frac{\omega}{\omega_{\rm Kep}}
\simeq 0.03
f(e) 
\left( {\frac {M_{\rm {env}}}{M_g}} \right) 
\left( {\frac {M_{\rm {env}}}{M_\odot}} \right)^ {-1/3} 
\left( {\frac {a} {10R_g}} \right)^ {-15/2}
\left( \frac {\dot M_w} {10^{-6} M_\odot \yr^{-1}} \right)^{-1},
\end{eqnarray}
where $f(e)\equiv f_s(e) (1-e^2)^{1/2}$ is a steeply increasing function
of $e$, with $f(0) \simeq 1$. 
I emphasize again that this relation holds only if the
time spent by the mass losing star in the giant phase is
shorter than the synchronization time $\tau_{\rm {syn}}$, only
for mass loss via a wind (not for a Roche lobe overflow), and
only if it gives $\omega<\omega_o$, where
$\omega_o$ is the orbital angular velocity.
The synchronization time is related to the circularization time
by the expression 
$\tau_{\rm {syn}} \simeq  (1+M_2/M_1)(M_2/M_1)^{-1}(I_g/M_gR_g^2) 
(R/a)^2 \tau_{\rm {cir}} $ (accurate for $e=0$), where $I_g$ is
the giant's moment of inertia.
Using the typical values as in equation (3), I find
equation (7) to hold for $3 R_g \lesssim a \lesssim 20 R_g$,
for a circular orbit; the separation can be larger for
eccentric orbits.

 Crudely, equation (7) indicates that in all symbiotic
systems with orbital period of
$P_{\rm orb} \lesssim 100 \yr$, with larger periods for larger
eccentricity, tidal interaction overcomes angular momentum loss,
and the systems are synchronized, or close to being so.
For systems with $a \gtrsim 5 R_g$, the situation is reversed during
the superwind phase at the termination of the AGB, when the mass
loss rate can be as large as $\sim 10^{-4} M_\odot \yr^{-1}$, and
the giant envelope loses synchronization and slows down.
This is more relevant to the formation of some bipolar planetary nebulae
than to SSs.

\subsection{Backflowing material}

The basic process here is that the hot companion prevents,
directly or indirectly, part of the mass blown by the giant
companion to acquire the escape velocity from the binary system.
That fraction of mass may acquire orbital angular momentum, and if
it is accreted back by the giant, it spins-up the giant's envelope.
 This process should be worked out in detail via numerical simulations,
e.g., similar to those of Mastrodemos \& Morris (1999),
but with more processes included, e.g., locally enhanced mass loss rate.
The conditions for a backflow to occur were examined in
an earlier paper (Soker 2001). 
I found there that for a significant backflow to occur, there should
be a slow dense flow, such that the relation between the total mass
in the slow flow, $M_i$, and the solid angle it covers
$\Omega$, is given by  $M_i/(\Omega/4 \pi) \gtrsim 0.1 M_\odot$.
 The requirement for both high mass loss rate per unit solid angle
and a very slow wind, such that it can be decelerated and flow back,
probably requires close binary interaction.
 This process, therefore, requires that the companion be close and
 already some spin-up of the giant's envelope have occurred.
 Large magnetic cool spots (see next section) may then lead
to a very slow and a high mass loss rate above these spots.
Because the escape velocity from the binary system is larger than that
from the mass losing star, some of this material may be accreted
back by the giant (or the companion). The gravitational interaction
with the binary system may transfer some orbital angular momentum to
the backflowing mass, e.g., in the accretion column formed behind
the companion.
The backflowing material, then, may possess large
specific angular momentum and spin-up the giant.

I suggest this process to account for the rotation of the SiO maser
shell in R Aquarii (Hollis et al.\ 2001).
Hollis et al. (2001) argue for a rotational period range of 8-34 yr,
at a radius of $\sim 1-2R_g$, where $R_g=1.8 \AU$ is the giant's radius.
The orbital period of the R Aquarii binary system is $\sim 44 \yr$, with
a semimajor axis of $\sim 17 \AU$, and an eccentricity of $\sim 0.8$
(Hollis et al.\ 2001).
The tidal synchronization time is very short for this system, hence,
the angular velocity of the spinning giant should have been
$\sim 44 \yr$.
 Any mass lost by the giant should have a longer rotation
period, and not as short as $\sim 8 - 34 \yr$.
The fast rotating maser spots are formed by backflowing material,
as is indeed observed by Boboltz, Diamond, \& Kamball (1997).
In particular, a high mass loss rate is expected during the
periastron passage.
A substantially  enhanced mass loss rate during periastron
passage may account for the high eccentricity of the system,
despite the short circularization time (Soker 2000a). 
 The backflowing mass, which is the source of the SiO maser,
further spins-up the giant's envelope.
Also, because of the fast rotation of the backflowing gas, there
will be a shear as it reaches the stellar surface.
Such a shear may further amplify the magnetic field near the
equator (Soker 2000b). 

\section{MAGNETIC ACTIVITY} 

The three processes discussed in the previous section show
that the giant companion in many SSs is likely to rotate much faster
than isolated, or in wide binary systems, RGB and AGB stars.
In the present section I discuss plausible implications of this
(relatively) fast rotation to the magnetic activity of these
cool giants.

\subsection{Global mass loss geometry}

A strong magnetic activity may influence the mass loss geometry,
e.g., cause a higher or lower mass loss rate in the equatorial plane,
and/or cause semi-periodic variation via a solar-like
magnetic activity cycle. 
The magnetic field may influence the mass loss process via dynamical
effects, i.e., the magnetic force due to pressure gradient
and/or tension directly influences the outflowing mass
(e.g., Chevalier \& Luo 1994;
Pascoli 1997; Matt et al.\ 2000; Blackman et al.\ 2001;
Garc\'{\i}a-Segura et al.\ 1999; Garc\'{\i}a-Segura, L\'opez,
\& Franco 2001; Gardiner \& Frank 2001).
As far as the shaping of circumstellar gas is concerned, my view is
(Soker \& Zoabi 2002; Soker 2002) that the global shaping mechanisms
are based on direct influence of a companion
(e.g., Mastrodemos \& Morris 1999) or indirect effects of
the magnetic field but not on direct dynamical effects of the
giant's magnetic field.
I find this to be the case in the SS system R Aquarii, from
the following consideration.
In the sun the magnetic activity determines the mass loss rate.
Indeed, the average solar X-ray luminosity resulting from the
magnetic activity is within an order of magnitude of the kinetic energy
carried by the solar wind.
In the SS R Aquarii, the X-ray luminosity of the central source is
much lower than the rate of kinetic energy carried by the giant's wind.
From the results of Kellogg, Pedelty, \& Lyon (2001), the central
X-ray luminosity is $L_x \sim 4 \times 10^{28} \erg \s^{-1}$.
 The rate of wind's kinetic energy is
$L_w = 3 \times 10^{30} \erg \s^{-1}
{(\dot M_w/10^{-7} M_\odot \yr^{-1})}(v_w/10 \km \s^{-1})^2$,
where $v_w$ is the wind speed.
The mass loss rate from R Aquarii is somewhat smaller than
the scaling above (see data summarized by Iben \& Tutukov 1996),
but the outflow speed is larger (Hollis et al.\ 1999).
The low ratio of $L_x/L_w < 0.01$ in R Aquarii hints that
the mass loss process from the cool giant is not determined
directly by magnetic activity.
Indirect effects of the magnetic field are very likely, though.
In $\S 3.3$ I speculate that some of the central X-emission
of R Aquarii may result from magnetic activity on the
giant's surface. 

An indirect effect can be the formation of magnetic cool spots,
above which dust formation, hence mass
loss rate, is enhanced (Soker 2000b and references therein).
Globally, the magnetic energy is much below that carried by the wind,
although in local spots the magnetic field can be strong.
Magnetic cool spots can be concentrated in the equatorial plane as
in the sun, leading to higher mass loss rate in the equatorial plane.
 It is also possible, probably in rapidly rotating stars,
that the spots are concentrated in the polar regions, as in rapidly
rotating MS stars (e.g., Schrijver \& Title 2001).
To conclude, relatively strong magnetic activity may lead to axially
symmetric mass loss process.

If a magnetic activity cycle exists, then the mass loss rate
may be semi-periodic (Soker 2000b). This means that a mass loss rate
presently observed in a specific system may not represent the
average mass loss rate. 

\subsection{Stochastic mass loss}

 One manifestation of magnetic cool spots is locally
enhanced mass loss rate, i.e., in particular directions the
mass loss rate is higher for a short time.
A similar effect can be caused by spots formed by
large convective elements (Schwarzschild 1975).
This means, as mentioned above, that the surface of the cool giant
can be asymmetric.
Another implication is a stochastic mass loss rate in the direction of
the hot companion, hence the accretion rate and geometry onto the
hot companion may vary during short periods, i.e., weeks to years
(much shorter than a possible magnetic cycle).
This stochastic accretion process may have the following
observable manifestations.
(1) Impulsive jets: A high mass flux in a specific direction toward the
companion, but not exactly on the line to it, means
that the flow has a high specific angular momentum about
the hot companion. If a disk was not present before, a temporary disk
may be formed. Such a disk may blow two jets for a short time.
(2) Density variation along the jets' axes:
 If a disk was present with two jets, the enhanced accretion
may lead to a denser jet. The consequence is two continuous jets,
but with variation in density along the jet axis. 
(3) A tilted and precessing disk:
 The enhanced mass flux may be away from the equatorial plane.
In such a case, the angular momentum accreted may have a direction
tilted with respect to the permanent accretion disk.
This may tilt the disk and cause a precession for a short time.
This may be observed as precessing jets, but for a limited
time.

Variation in the mass accretion rate may result from the
pulsation of the giant as well, but to a lesser extent than
I expect in a strongly magnetically active giant. 

\subsection{X-ray emission}

Magnetically active MS stars have X-ray luminosity of
up to $\sim 10^{30} \erg \s^{-1}$ (Stelzer \& Neuh\"auser 2001).
 The source of the X-ray luminosity is a hot corona
and magnetic reconnection events, e.g., flares.
Some observations indicate that cool giants may also emit
in the X-ray band (H\"unsch et al.\ 1998;
Schr\"oder, H\"unsch \& Schmitt 1998; H\"unsch 2001),
resulting most likely from magnetic activity.
The presence of magnetic fields in AGB stars is inferred from
maser polarization (e.g., Kemball \& Diamond  1997). 
It seems that even a very slow rotation is enough to
sustain a turbulent dynamo activity in AGB stars,
i.e., an $\alpha^2 \omega$ dynamo (Soker \& Zoabi 2002).
The fast rotating cool giants in SSs may be among the most
magnetically active cool giants and those in which strong X-ray
emission is expected.
There is no basic dynamo theory to predict the level of magnetic
activity in AGB stars; crude estimates for AGB stars rotating
with periods of $\sim 10^2-10^3 \yr$ give
$\dot E_B < 10^{30}-10^{32} \erg \s^{-1}$
(Soker \& Harpaz 1992).
 The temperature of the X-ray emitting plasma is most likely
$T_x \sim 10^6 \K$, as in MS stars.
Magnetic flares may lead to detectable X-ray variability on
time scales of weeks to months.
 I therefore speculate that the X-ray central source in the SS
R Aquarii, with  $L_x \simeq 4 \times 10^{28} \erg \s^{-1}$
and $T_x=2 \times 10^6 \K$ (Kellogg et al.\ 2001)
result from a magnetic activity on the surface of the Mira variable.
X-ray observations of other SSs is highly encourage.
 
\section{SUMMARY} 

The main goal of the present paper is to point out that the
cool giant companion in most symbiotic binary systems (SSs) are
likely to be strongly (relative to other cool giants) magnetically
active. 

The observational aspects of the magnetic activity are:
(1) Imposing a global axisymmetrical mass loss geometry from the
cool giant, i.e., a higher mass loss rate in the equatorial or
polar directions.
(2) Stochastic mass loss process, i.e., variation of mass loss rate
with time and location on the giant's surface.
The hot companion itself may cause a much larger departure from
spherical symmetry in the circumbinary matter.
Therefore, the asymmetry due to magnetic activity, both
global and local, should be looked for near the giant's surface,
e.g., SiO maser spots.
The stochastic mass loss process may also cause variations in the
intensity and direction of jets blown by an accreting companion.
(3) As in main sequence stars, the magnetic activity may
lead to X-ray emission.
The expected temperature is $\sim 10^6 \K$.
The X-ray luminosity is hard to predict as there is no
basic dynamo model for cool giants.
 A crude estimate is $L_x < 10^{30}-10^{32} \erg \s^{-1}$,
in any case below the kinetic energy of the giant's wind.
I suggest that this may contribute to the central X-ray source in
R Aquarii, and hence I encourage more X-ray observations of SSs.

In arguing for magnetic activity of the cool giants of SSs,
I assumed, based on the behavior of main sequence stars,
that magnetic activity and X-ray luminosity increase
with rotation velocity.
I then showed that the cool companions in SSs are likely
to rotate much faster than isolated, or in wide binary systems,
cool giants, because of three processes:
(1) In SSs in which the hot companion is a WD, the presently
giant accreted mass from the wind of the WD's progenitor
during the AGB phase of this progenitor. The accreted mass has
a large specific angular momentum, hence it spins-up the MS progenitor.
(2) The giant is spun up via tidal interaction with the
hot companion.
(3) Some fraction of the mass blown by the giant may
flow back to the giant after acquiring angular momentum from
the binary system, e.g., gravitational interaction with the
companion.     

\acknowledgements        
 This research was supported by the US-Israel
Binational Science Foundation.

\end{document}